\documentclass[
 aip,
 jcp,
 amsmath,amssymb,
 reprint,
]{revtex4-1}

\usepackage{graphicx}
\usepackage{dcolumn}
\usepackage{bm}
\usepackage[utf8]{inputenc}
\usepackage[T1]{fontenc}
\usepackage{mathptmx}
\usepackage{etoolbox}
\usepackage{mathtools}
\usepackage{mathrsfs}
\usepackage{amsthm}
\usepackage{physics}
\usepackage{xcolor}
\usepackage{enumitem}
\usepackage{microtype}
\usepackage{hyperref}

\allowdisplaybreaks

\makeatletter
\def\@email#1#2{%
 \endgroup
 \patchcmd{\titleblock@produce}
  {\frontmatter@RRAPformat}
  {\frontmatter@RRAPformat{\produce@RRAP{*#1\href{mailto:#2}{#2}}}\frontmatter@RRAPformat}
  {}{}
}%
\makeatother

\newcommand{\R}{\mathbb{R}}

\newcommand{\Flieb}{F}

\newcommand{\pair}[2]{\langle #1, #2\rangle}
\newcommand{\Trf}{\operatorname{Tr}}

\newcommand{\Xrho}{X_\rho}

\newcommand{\EBO}{E_{\mathrm{BO}}}
\newcommand{\Vnn}{V_{\mathrm{nn}}}
\newcommand{\FF}{\mathcal F}
\newcommand{\RR}{\mathcal R}
\newcommand{\VV}{\mathcal V}
\newcommand{\chiR}{\chi_R}
\newcommand{\rhoR}{\rho_R}
\newcommand{\dr}{\,d\mathbf r}
\newcommand{\drr}{\,d\mathbf r\,d\mathbf r'}
\newcommand{\Phimap}{\Phi}
\newcommand{\Xv}{X_v}

\begin{document}

\preprint{AIP/123-QED}

\title{A density-functional perspective on force fields}

\author{Nan Sheng}
\affiliation{
Institute for Computational and Mathematical Engineering (ICME),
Stanford University, Stanford, CA 94305, USA.
}
\email{nansheng@stanford.edu}

\date{\today}

\begin{abstract}
Force fields are usually formulated directly in nuclear configuration space, whereas density functional theory is naturally formulated in terms of external potentials, densities, and variational duality. We show that exact force fields are variationally induced by DFT: the Born--Oppenheimer potential-energy surface is the pullback of the external-potential energy functional along the map from nuclear configurations to Coulomb potentials. In the Lieb formulation of density functional theory, the density is the first functional derivative of the energy with respect to the external potential, while the density--density response function is the second. Pulling these derivative objects back to nuclear configuration space yields the force and the nuclear Hessian, together with explicit terms induced by the nuclear-generated potential and the nuclear--nuclear repulsion. The resulting picture places force fields, density functional theory, and response theory within a single derivative hierarchy. The purpose of the present work is conceptual rather than algorithmic.
\end{abstract}

\newtheorem{definition}{Definition}
\newtheorem{proposition}{Proposition}

\maketitle

\section{Introduction}

Force fields are among the oldest and most widely used objects in atomistic modeling. In their classical form, they assign an energy to a nuclear configuration through prescribed functional forms built from bonded and nonbonded interactions, thereby defining the forces governing molecular dynamics, structural relaxation, and vibrational behavior.\cite{FrenkelSmit2023} Over time, this notion has broadened considerably. Alongside traditional molecular-mechanics models, one now has increasingly accurate interatomic potentials derived from electronic-structure data, as well as modern flexible surrogates learned from such data.\cite{Behler2016,Deringer2019,Mishin2021,Muser2023} Yet across these different realizations, the primitive viewpoint remains the same: a force field is treated as a scalar function on nuclear configuration space.

Electronic-structure theory enters this picture through the Born--Oppenheimer approximation. For fixed nuclei, one solves an electronic ground-state problem and defines the nuclear potential-energy surface from the resulting electronic energy together with the nuclear--nuclear repulsion. In density functional theory (DFT), this electronic problem is naturally formulated in terms of the external potential, the electron density, and their variational duality.\cite{HohenbergKohn1964,KohnSham1965,Lieb1983,Jones2015,Burke2012} In practice, however, these two viewpoints are usually presented in rather different languages. Force fields are introduced as coordinate-space energy models, while DFT is introduced as an electronic variational theory. The relation between them is of course physically understood, but it is less often isolated as a formal organizing principle in its own right.

Recent work has begun to narrow this gap from several directions, including data-driven maps between external potentials, densities, and energies, as well as external-potential-centered atomistic surrogates.\cite{Brockherde2017,Shao2023,Das2025FTA} Morante and Rossi used the Legendre structure of the ground-state energy as a functional of the external potential to derive the atomic-force expression required in Born--Oppenheimer molecular dynamics.\cite{MoranteRossi2017} The present work addresses a different, structural question: the relation between the full Born--Oppenheimer potential-energy surface and the external-potential formulation of DFT.

The purpose of the present work is to formulate this structural relation explicitly and to develop its consequences. We show that exact force fields are variationally induced by DFT: the Born--Oppenheimer potential-energy surface may be written as
\begin{equation}
\EBO(R)=E[v_R]+\Vnn(R),
\label{eq:intro_main}
\end{equation}
so that the force field is the pullback of the external-potential energy functional along the map from nuclear configurations to Coulomb potentials. From this perspective, the density is the first derivative object in external-potential space, the density--density response function is the second, and the force and nuclear Hessian are the corresponding induced objects in nuclear configuration space. The resulting picture places force fields, density functional theory, and response theory within a single derivative hierarchy. The aim of the present article is not to propose a new numerical force-field scheme, but to provide a compact variational and density-functional framework for understanding what a force field is.

\section{Force fields as scalar fields on nuclear configuration space}

We begin by fixing the object to be interpreted.

Let $\RR$ denote the nuclear configuration space for a system with fixed nuclear charges $\{Z_I\}$. Abstractly, an element $R \in \RR$ specifies the allowed nuclear positions $\{R_I\}$. A force field is then simply a scalar-valued function on $\RR$.

\begin{definition}
A force field is a map
\begin{equation}
\FF : \RR \to \R,
\end{equation}
which assigns to each nuclear configuration $R \in \RR$ an energy $\FF(R)$.
\end{definition}

In most applications, the force field is assumed to be differentiable away from singular configurations. Its first derivative gives the force and its second derivative gives the Hessian:
\begin{equation}
F_I(R) = -\frac{\partial \FF(R)}{\partial R_I},
\qquad
H_{IJ}(R)=\frac{\partial^2 \FF(R)}{\partial R_I\partial R_J}.
\label{eq:force_hessian_def}
\end{equation}
Higher derivatives encode increasingly refined local structure of the energy surface, including anharmonic couplings and higher-order nuclear response.

At this level, a force field is simply a scalar field on nuclear configuration space. This definition is deliberately minimal. It does not presuppose a classical functional form, an \textit{ab initio} construction, or any particular approximation scheme. The physically relevant question is how such a scalar field arises from the electronic many-body problem.

Within the Born--Oppenheimer framework, the physically realized force field is the ground-state Born--Oppenheimer potential-energy surface
\begin{equation}
\FF(R)=\EBO(R),
\end{equation}
obtained from the electronic ground-state problem at fixed nuclei plus the explicit nuclear repulsion. Our goal is to reinterpret this object in a way natural from the standpoint of density functional theory.

\section{From nuclear configurations to external potentials}

A nuclear configuration determines an external potential. For fixed nuclear charges
$\{Z_I\}_{I=1}^M$ at positions $\{R_I\}_{I=1}^M$, the corresponding point-nuclear
Coulomb potential is
\begin{equation}
v_R(\mathbf r)
=
-\sum_{I=1}^M \frac{Z_I}{\abs{\mathbf r-R_I}}.
\label{eq:vR}
\end{equation}
This defines a map
\begin{equation}
\Phimap : \RR \to \VV,
\qquad
\Phimap(R)=v_R,
\label{eq:Phi_map}
\end{equation}
where $\VV$ denotes a suitable external-potential space.

The image of $\Phimap$ is, of course, not the full external-potential space, but only
the physically realized class of nuclear-generated Coulomb potentials. After excluding nuclear collisions, the map descends to an injective map on nuclear configuration space modulo permutations of identical nuclei: the singularity set of $v_R$ recovers the nuclear positions, while the corresponding singular coefficients recover the nuclear charges. Thus passing from $R$ to $v_R$ does not discard physical information; it only changes the variable space in which the same nuclear configuration is represented.

\section{Density-functional formulation and duality}

We now formulate the electronic ground-state problem in the density-functional language.
Let $\Xrho$ denote a density space and $\Xv=\Xrho^*$ the dual potential space, with canonical pairing
\begin{equation}
\pair{v}{\rho}
=
\int v(\mathbf r)\rho(\mathbf r)\,\dr.
\label{eq:pairing}
\end{equation}
For definiteness, one may have in mind the standard Lieb setting,\cite{Lieb1983}
\begin{equation}
\Xrho = L^1(\Omega)\cap L^3(\Omega),
\qquad
\Xv = L^\infty(\Omega)+L^{3/2}(\Omega),
\label{eq:spaces}
\end{equation}
although the discussion below is structural and does not depend on a full function-space analysis.

The exact interacting universal functional is
\begin{equation}
\Flieb[\rho]
:=
\inf_{\Gamma\mapsto\rho}
\Trf\!\bigl(\Gamma(\hat T+\hat W)\bigr).
\label{eq:Lieb}
\end{equation}
The associated source-side energy functional and its Legendre--Fenchel dual relation are
\begin{equation}
\left\{
\begin{aligned}
E[v]
&=
\inf_{\rho}
\left\{
\Flieb[\rho]+\pair{v}{\rho}
\right\},
\\[2pt]
\Flieb[\rho]
&=
\sup_{v}
\left\{
E[v]-\pair{v}{\rho}
\right\}.
\end{aligned}
\right.
\label{eq:duality_pair}
\end{equation}
Thus $\rho$ and $v$ are dual variables, while $E[v]$ and $\Flieb[\rho]$ are the corresponding dual variational objects.~\cite{Lieb1983,ShengExactDFT2026,ShengInverseKS2026} In the general convex-analytic setting, their relation may be written as
\begin{equation}
\rho \in \partial^{+}E[v]
\iff
-\,v \in \partial \Flieb[\rho].
\label{eq:subgradient_pair}
\end{equation}
In the regular differentiable regime considered in the present work, this reduces to
\begin{equation}
\left\{
\begin{aligned}
\frac{\delta E[v]}{\delta v(\mathbf r)}
&=
\rho_v(\mathbf r),
\\[2pt]
\frac{\delta \Flieb[\rho]}{\delta \rho(\mathbf r)}
&=
-\,v(\mathbf r).
\end{aligned}
\right.
\label{eq:duality_differential_pair}
\end{equation}
Throughout, if \(A_v\) denotes a \(v\)-dependent quantity on external-potential space, we write \(A_R\) for its pullback along \(R\mapsto v_R\), namely \(A_R:=A_{v_R}\).

Given a nuclear configuration $R$, the Born--Oppenheimer energy surface is
\begin{equation}
\EBO(R)=E[v_R]+\Vnn(R),
\label{eq:EBO_pullback}
\end{equation}
with
\begin{equation}
\Vnn(R)=\sum_{I<J}\frac{Z_I Z_J}{\abs{R_I-R_J}}.
\label{eq:Vnn}
\end{equation}
Equivalently,
\begin{equation}
\EBO=\Phimap^*E+\Vnn.
\label{eq:pullback}
\end{equation}
This is the basic structural relation of the present work: the force field in nuclear
configuration space is the pullback of the external-potential energy functional, together
with the explicit nuclear term.

\section{First-order structure: density and force}

At zeroth order, the force field itself is simply the pulled-back scalar value in
Eq.~\eqref{eq:EBO_pullback}. The first-order structure is obtained by differentiating this
relation with respect to the nuclear coordinates.

The first member of Eq.~\eqref{eq:duality_differential_pair} identifies the density as the
first derivative object naturally attached to the external-potential energy functional.
Evaluating this relation at \(v=v_R\), the first-order structure is obtained by differentiating
Eq.~\eqref{eq:EBO_pullback} with respect to a nuclear coordinate \(R_I\):
\begin{equation}
\frac{\partial \EBO(R)}{\partial R_I}
=
\int \rhoR(\mathbf r)
\frac{\partial v_R(\mathbf r)}{\partial R_I}\,\dr
+
\frac{\partial \Vnn(R)}{\partial R_I},
\label{eq:Eprime}
\end{equation}
where $\rhoR$ is the electronic ground-state density corresponding to the nuclear
configuration $R$. Therefore the nuclear force is
\begin{equation}
F_I(R)
=
-
\int \rhoR(\mathbf r)
\frac{\partial v_R(\mathbf r)}{\partial R_I}\,\dr
-
\frac{\partial \Vnn(R)}{\partial R_I}.
\label{eq:force_final}
\end{equation}

The atomic-force expression itself is familiar and was previously derived from the Legendre structure of the external-potential energy functional by Morante and Rossi in the context of Born--Oppenheimer molecular dynamics.\cite{MoranteRossi2017} In the present formulation, it arises as the first-order consequence of the pullback relation for the full Born--Oppenheimer potential-energy surface. The density is not itself the force. Rather, the density is the gradient of the energy in external-potential space, and the nuclear force is the induced first-order object obtained by composing that gradient with the geometry of the map $R \mapsto v_R$.

Thus the first-order structure of a force field is inherited from the first-order variational structure of density functional theory.

\section{Second-order structure: response and Hessian}

The second-order structure is closely related to the objects computed in
density-functional perturbation theory, where electronic linear response
is used to obtain phonons, interatomic force constants, dielectric
response, and related properties.\cite{Baroni2001DFPT}
The emphasis here is different. Rather than developing a perturbative
algorithm, we identify the pullback relation that makes the nuclear
Hessian the coordinate-space descendant of the external-potential
response kernel.

In the same regular setting, the second functional derivative of the energy with respect to the external potential defines the density--density response function, which is the Hessian of the external-potential energy functional:
\begin{equation}
\chi_v(\mathbf r,\mathbf r')
=
\frac{\delta \rho_v(\mathbf r)}{\delta v(\mathbf r')}
=
\frac{\delta^2 E[v]}{\delta v(\mathbf r)\delta v(\mathbf r')}.
\label{eq:chi}
\end{equation}

We now differentiate Eq.~\eqref{eq:Eprime} once more:
\begin{align}
\frac{\partial^2 \EBO(R)}{\partial R_I\partial R_J}
&=
\int
\frac{\partial \rhoR(\mathbf r)}{\partial R_J}
\frac{\partial v_R(\mathbf r)}{\partial R_I}\,\dr
\nonumber\\
&\quad+
\int
\rhoR(\mathbf r)
\frac{\partial^2 v_R(\mathbf r)}{\partial R_I\partial R_J}\,\dr
+
\frac{\partial^2 \Vnn(R)}{\partial R_I\partial R_J}.
\label{eq:hessian_step1}
\end{align}
Applying the chain rule again gives
\begin{equation}
\frac{\partial \rhoR(\mathbf r)}{\partial R_J}
=
\int
\chiR(\mathbf r,\mathbf r')
\frac{\partial v_R(\mathbf r')}{\partial R_J}\,d\mathbf r'.
\label{eq:density_chain}
\end{equation}
Substituting this into Eq.~\eqref{eq:hessian_step1}, we obtain
\begin{align}
\frac{\partial^2 \EBO(R)}{\partial R_I\partial R_J}
&=
\iint
\chiR(\mathbf r,\mathbf r')
\frac{\partial v_R(\mathbf r)}{\partial R_I}
\frac{\partial v_R(\mathbf r')}{\partial R_J}\,\drr
\nonumber\\
&\quad+
\int
\rhoR(\mathbf r)
\frac{\partial^2 v_R(\mathbf r)}{\partial R_I\partial R_J}\,\dr
+
\frac{\partial^2 \Vnn(R)}{\partial R_I\partial R_J}.
\label{eq:hessian_final}
\end{align}

Equation~\eqref{eq:hessian_final} is the second-order counterpart of the force formula. It shows that the nuclear Hessian is not identical to the response function. Rather, it is induced from the response function together with an explicit curvature term coming from the second derivative of the nuclear-generated external potential and the explicit nuclear--nuclear term.

This distinction between the potential-space Hessian and the nuclear-space Hessian is essential. The former is the exact two-point response object associated with the external-potential functional, while the latter is the induced coordinate-space second derivative obtained after pullback. In this sense, the response kernel is the more primitive second-order electronic object, and the force-constant matrix is its induced nuclear counterpart.

\section{Higher-order structure}

The pattern above extends formally to higher order. Whenever higher functional derivatives of $E[v]$ exist, they define higher-order response objects in external-potential space. Repeated differentiation of the pullback relation
\begin{equation}
\EBO=\Phimap^*E+\Vnn
\end{equation}
then yields higher derivatives of the nuclear energy surface through repeated application of the chain rule.

Thus the higher-order derivatives of the force field are built from two ingredients:
\begin{enumerate}[leftmargin=*,itemsep=1pt]
\item the higher-order derivative hierarchy of the external-potential functional $E[v]$, and
\item the higher-order geometry of the map $\Phimap:R\mapsto v_R$.
\end{enumerate}
In this sense, the force field on nuclear configuration space is not an isolated scalar function devoid of further structure. It is the image of an entire derivative tower defined on external-potential space.

This viewpoint suggests that energy, density, force, Hessian, and higher-order response should be viewed not as unrelated objects, but as successive levels in a single variational hierarchy.

\section{Discussion}

The formulation developed above suggests a stronger conclusion than a mere change of notation. In the Born--Oppenheimer setting, force fields are not external to DFT. Their more fundamental status is variational: the usual energy surface in nuclear configuration space is the pullback of the external-potential energy functional to nuclear configuration space.

From this perspective, force calculations and response calculations are not conceptually separate tasks. The density is the first derivative object of the external-potential energy functional, and the density--density response function is the second. The nuclear force and nuclear Hessian are the corresponding induced first- and second-order objects in nuclear configuration space. Thus the usual coordinate-space hierarchy of energy, force, and Hessian may be read as the pulled-back image of a higher derivative hierarchy in external-potential space.

This viewpoint also clarifies the status of different styles of force-field approximation. Ordinary force-field models, whether classical, \textit{ab initio}-derived, or data-driven, may be understood as approximations to the pulled-back object $\EBO(R)$. More structured models may instead attempt to approximate, explicitly or implicitly, higher-level objects such as $E[v]$, $\rho_v$, or $\chi_v$. The present work does not advocate any particular numerical choice. Its point is structural: these objects belong to one variational hierarchy and should not be regarded as unrelated observables.

The same hierarchy suggests a possible source of structured supervision for data-driven force fields. Such models are commonly trained on energies and forces expressed directly in nuclear configuration space.\cite{Behler2016,Deringer2019,Mishin2021,Muser2023} By contrast, $\rho_v$ and $\chi_v$ encode the first- and second-order derivative structure of the energy in external-potential space, from which the electronic contributions to nuclear forces and Hessians are induced by pullback. Density and response data may therefore provide auxiliary learning targets that constrain not only the pulled-back potential-energy surface, but also the ambient variational structure that generates its coordinate-space derivatives. The practical value of such supervision remains an algorithmic question, especially because the full response kernel is high-dimensional and generally nonlocal.

\section{Scope and limitations}

The present work is conceptual in scope. It does not propose a new practical force-field construction, nor does it claim that external-potential space is the most efficient variable space for numerical modeling. In continuous form, the potential-space formulation naturally introduces spatial integrals and nonlocal response kernels that are typically much more cumbersome than direct coordinate-space representations.

Several analytical issues have also been left at the structural level. In a fully rigorous treatment, one must distinguish carefully between the effective domain of $\Flieb$, the representability class admitting supporting potentials, and the regular regime in which the functional derivatives of $E[v]$ may be treated classically. In nonsmooth settings, the appropriate language is that of subgradients and supporting potentials rather than ordinary derivatives. Likewise, the discussion of the map $R \mapsto v_R$ has been restricted to collision-free point-nuclear Coulomb configurations, for which the map descends to an injective representation of nuclear configurations modulo permutations of identical nuclei.

Finally, the discussion has been confined to the ground-state Born--Oppenheimer setting. Finite-temperature, grand-canonical, time-dependent, and reduced-density-matrix generalizations are all possible, and in some respects may offer even cleaner dual formulations. These directions, however, lie beyond the scope of the present article.

\section{Conclusion}

We have shown that exact force fields are variationally induced by DFT. The central observation is that the Born--Oppenheimer potential-energy surface is the pullback of the external-potential energy functional along the map from nuclear configurations to Coulomb potentials. Within the Lieb formulation of density functional theory, this places the density and the density--density response function as the first and second functional derivatives of the external-potential energy.

Pulling these derivative objects back to nuclear configuration space yields the force and the nuclear Hessian, together with the explicit terms induced by the nuclear-generated potential and the nuclear--nuclear repulsion. The resulting picture places force fields, density functional theory, and response theory within a single derivative hierarchy. In this sense, a force field in the Born--Oppenheimer setting is not merely a scalar function of nuclear coordinates, but the pulled-back coordinate-space image of a higher-level variational object.

\begin{acknowledgments}
The author thanks Yuzhi Liu for helpful discussions.
\end{acknowledgments}

\bibliography{ref}

\end{document}